# Superconductivity in Metal-Rich Chalcogenide Ta$_2$Se

Xin Gui,[1] Karolina Górnicka,[2] Tomasz Klimczuk,[2] Weiwei Xie[1]*

[1] Department of Chemistry, Louisiana State University, Baton Rouge, LA, USA 70803

[2] Faculty of Applied Physics and Mathematics, Gdansk University of Technology, Narutowicza 11/12, Gdansk, Poland 80–233

Address correspondence to E-mail: weiweix@lsu.edu; weiwei.xie@rutgers.edu

Phone: 225-578-1074

***ABSTRACT***

The metal-metal bond in metal-rich chalcogenide is known to exhibit various structures and dominate interesting physical properties. Ta$_2$Se can be obtained by both arc-melting and solid-state pellet methods. Ta$_2$Se crystallizes a layered tetragonal structure with space group *P*4/*nmm* (S.G.129, Pearson symbol *tP*6). Each unit cell consists of four layers of body-centered closed packing Ta atoms sandwiched between two square nets of Se atoms, forming the Se-Ta-Ta-Ta-Ta-Se networks. A combined result of magnetic susceptibility, resistivity, and heat capacity measurements on Ta$_2$Se indicate the bulk superconductivity with $T_c$ = 3.8 (1) K. According to the first-principal calculations, the *d* orbitals in Ta atoms dominate the Fermi level in Ta$_2$Se. The flat bands at Γ-point in the Brillouin zone (BZ) yield to the van Hove singularities in density of states (DOS) around the Fermi level, which is intensified by introducing spin-orbit coupling (SOC) effect, thus, could be critical for the superconductivity in Ta$_2$Se. The physical properties especially superconductivity is completely different from Ta-rich alloys or transition metal dichalcogenide TaSe$_2$.

Transition metal-rich chalcogenides are a fascinating series of solid-state structures in which several-atom-thick slabs of transition metal atoms are terminated with monolayers of chalcogenides. Moreover, transition metal-rich chalcogenides usually share a common structural feature that transition metal forms octahedron or trigonal prism with chalcogen elements in the center,[1–4] which can be considered as the anti-format of transition metal dichalcogenides ($TM_2$).[5] The layered transition-metal dichalcogenides have received interests for decades due to the varieties in electronic properties in both bulk and surface states.[6–12] However, limited study of transition metal-rich chalcogenides, mainly with an emphasis on the structures, has been reported.[13–15] Most metal-rich chalcogenides occur in the earliest transition metals, Sc, Y, Ti, and especially the late lanthanides.[14–19] Harbrecht discovered the first of these transition metal-rich chalcogenides, $Ta_2Se$, by a simple arc melting preparation.[2] By comparing the crystal structures of $Ta_2Se$ and the well-known 2H-$TaSe_2$, one can obtain some significant similarities easily. The layered stacking pattern of 2H-$TaSe_2$ can be seen as (TaSe)SeSe(TaSe) while it is (TaSe)TaTa(TaSe) for $Ta_2Se$. The only difference is that the Ta atoms in $TaSe_2$ are three-coordinated but four-coordinated in $Ta_2Se$. Interestingly, when Cu atoms were intercalated into the $TaSe_2$ van der Waals gap, superconducting transition temperature can be increased from 0.14 K for pure 2H-$TaSe_2$ to $T_{c-max}$ = 2.7 K $Cu_xTaSe_2$.[20] Thus, for $Ta_2Se$, will the similar intercalation of Ta atoms induce the superconductivity and yield a high $T_c$?

The preparation of $Ta_2Se$ and phase determination method are shown in Supporting Information. Obtained $Ta_2Se$ chunk from arc-melting was determined to contain a small amount of 1T-$TaSe_2$ (P-3$m$1) as the impurity (~ 6.5 wt%). The powder X-ray diffraction (XRD) pattern shown in Figure 1$b$ matches with the previously reported $Ta_2Se$ structure very well.[2] As shown in Figure 1$a$, two crystallographically different Ta sites, marked as Ta1 and Ta2, and one Se site exist in $Ta_2Se$ binary compound. Specifically, Ta1 bilayer was sandwiched by two edge-shared Ta2@$Se_4$ layers and the resulted Se-Ta2-Ta1-Ta1-Ta2-Se layers are stacking along $c$-axis to form a layered $Ta_2Se$ structure. The Ta1-Ta1 and Ta1-Ta2 bonds are 2.831 (2) Å and 2.895 (2) Å, respectively while the Ta2 atoms are separating with Se with a length of 2.665 (3) Å. The long Se-Se distance (~3.57 Å) indicates the van der Waals force bonds the Se-$Ta_4$-Se layers in $Ta_2Se$. Moreover, the chemical composition was also determined by Scanning Electron Microscope-Energy Dispersive X-Ray Spectroscopy (SEM-EDS) shown in table S1 which indicates a formula of $Ta_{1.92(6)}Se$ and the excess Se shows that $TaSe_2$ can be a plausible impurity.

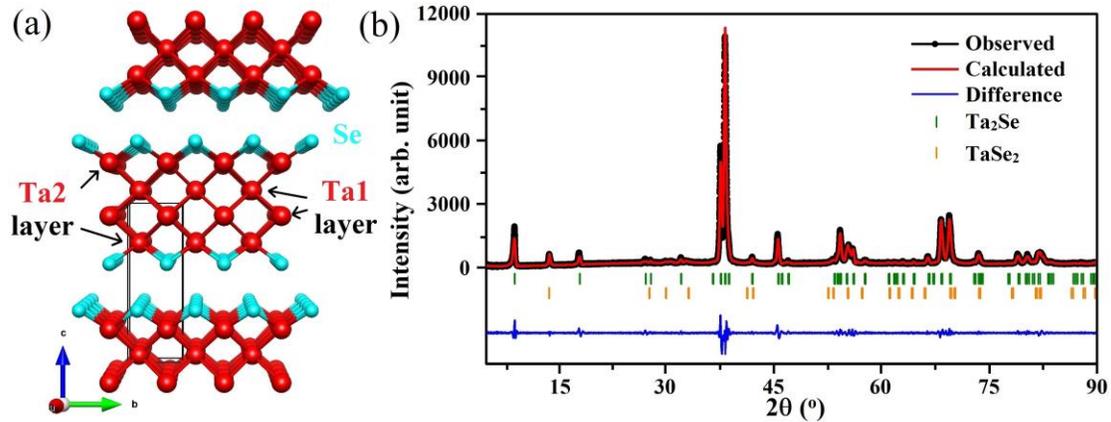

**Figure 1 (a).** The crystal structure of Ta$_2$Se where red and cyan balls represent Ta and Se atoms, respectively. **(b).** Refined powder XRD pattern for Ta$_2$Se. Black line with balls, red line, blue line, green vertical ticks and orange vertical ticks stand for observed, calculated patterns, difference between observed and calculated patterns, Ta$_2$Se Bragg peaks and 1T-TaSe$_2$ Bragg peaks.

Detailed physical properties measurements are described in Supporting Information. The plot of the volume magnetic susceptibility ($\chi_V$) versus temperature after a diamagnetic correction is shown in Figure 2a. The large diamagnetic signal below 3.8 K indicates the occurrence of superconductivity in this compound. Based on the ZFC signal, the transition is broad, likely due to the relatively large applied field (50 Oe) and does not saturate even at the lowest available temperature. However, $4\pi\chi_V$ (2K) = -1.07, which an absolute value is larger than the expected for the full Meissner fraction ($4\pi\chi_V$ = -1). This discrepancy is caused by a demagnetization effect and dependent on the sample shape and its orientation with respect to the direction of the external magnetic field. The FC signal is much weaker compared to the ZFC signal, usually resulting from strong flux trapping in Ta$_2$Se and is typically observed in polycrystalline samples. The critical superconducting temperature (T$_c$) was estimated as the intersection between two lines marked in red in Figure 2a: the first one is the steepest slope line of the superconducting signal and the second is an extrapolation of the normal metal state $\chi_V$ to lower temperature.[21] The value thus obtained is T$_c$ = 3.85 K, higher than the critical temperature for TaSe$_2$ (T$_c$ = 0.22 K[22] and lower than reported for pure Ta metal (T$_c$ = 4.4 K).[23] The width of the transition and the critical temperature decrease as increasing the applied magnetic field and the signal completely vanished above 3500 Oe. The field-dependence of magnetization is shown in Figure S2 which indicates a typical character of type-II superconductors.

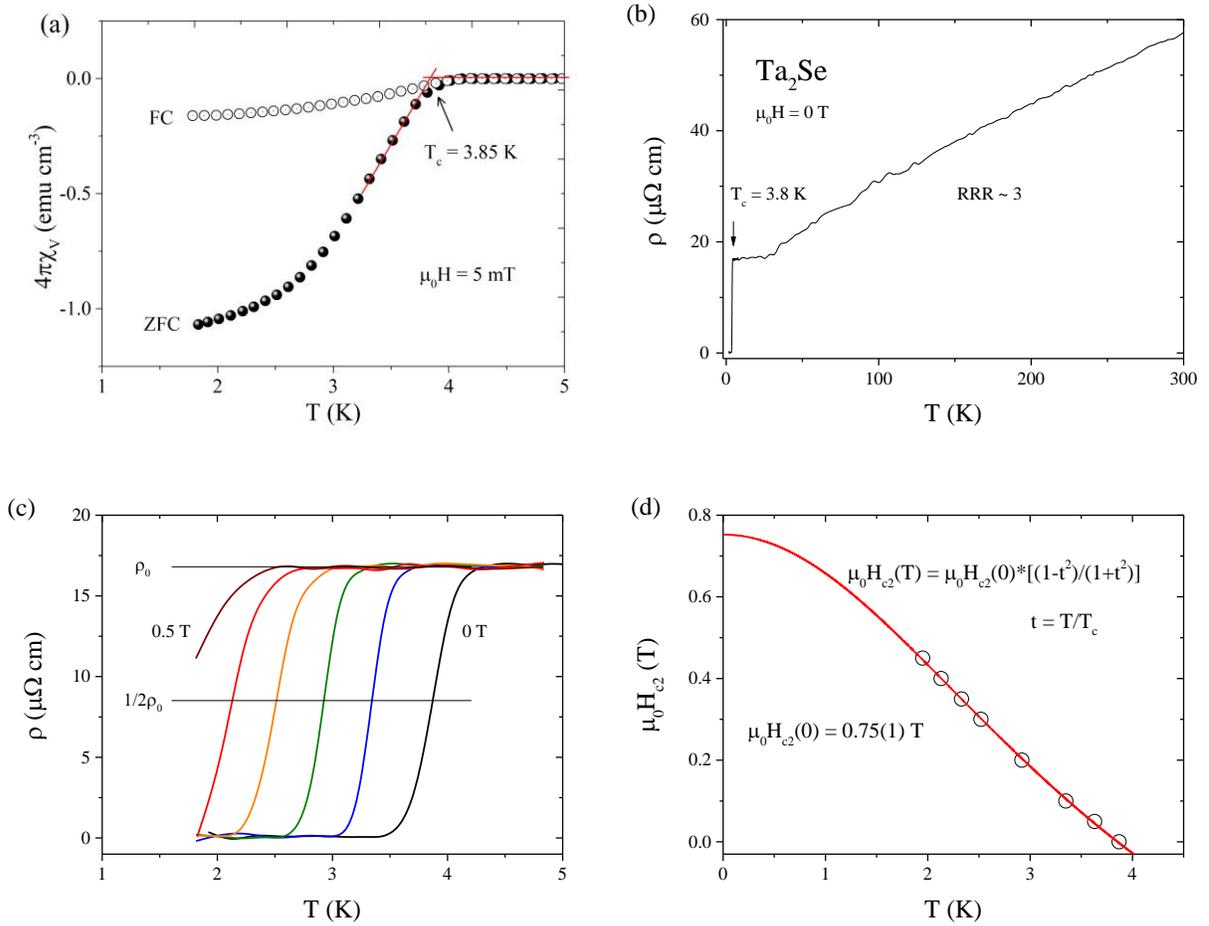

**Figure 2 (a).** The temperature-dependence of zero-field cooling (ZFC) and field-cooling volume magnetic susceptibility for Ta$_2$Se. The data were collected between 1.8 K to 5K in applied magnetic field $\mu_0H$ = 5 mT. **(b).** Electrical resistivity measured $\rho$(T) of Ta$_2$Se measured in zero magnetic field. **(c).** Expanded plot of the low-temperature $\rho$(T) showing the superconducting transition for different magnetic field from 0 T to 0.5 T. Horizontal lines represent a residual resistivity and a half of the transition, respectively. **(d).** Upper critical field $\mu_0H_{c2}$ vs. temperature of Ta$_2$Se determined from the electrical resistivity $\rho$(T,H) data in panel **(c)**. The red curve is a fit obtained by using a Ginzburg–Landau (G–L) equation.

Subsequently, the resistivity measurements were carried out in Quantum Design Dynacool with the four-probe technique. Figure 2*b* presents the resistivity as a function of temperature in the range of 1.8-300 K without applying external magnetic field. The resistivity undergoes a sudden drop at 3.8 K, which is an indication of superconductivity. In the normal state, the $\rho$(T) curve exhibits metallic behavior of a Bloch-Grüneisen type. Typical behaviors in resistivity for polycrystalline metals were observed with low residual resistivity ratio RRR ($\rho$(300 K)/$\rho$(4 K) = 3). Figure 2*c* emphasizes the low-temperature resistivity under various magnetic fields from 0 to

0.5 T. At $\mu_0H = 0T$, an abrupt resistivity drop due to the superconducting transition is clearly observed at $T_c = 3.8$ K. As can be seen, the superconducting transition temperatures were suppressed with larger fields. Above 1.8 K, the zero-resistance behavior is not observed for $\mu_0H = 0.5$ T and the resistivity drop disappears for $\mu_0H > 0.75$ T (not shown here). Using the criterion that the point with 50% normal state resistivity suppressed can be considered as the transition temperature, we determined the upper critical fields $\mu_0H_{c2}(T)$ for $Ta_2Se$ at various temperatures below 3.8 K (see Figure 2d). The data are fitted with the following Ginzburg-Landau relation[24]:

$$\mu_0H_{c2}(T) = \mu_0H_{c2}(0)\frac{(1-t^2)}{(1+t^2)}$$

where $t = T/T_c$ and $T_c$ is a fitting parameter (transition temperature at zero magnetic field). The G-L relation well described the experimental data and it yields $\mu_0H_{c2}(0) = 0.75(1)$ T and $T_c = 3.86(1)$ K. The obtained upper critical field is not exceeding the Pauli limiting field for the weak coupling BCS superconductors[25] $H_{c2}^p(0) = 1.85T_c$, which for $T_c = 3.8$ K gives $H_{c2}^p(0) = 7$ T.

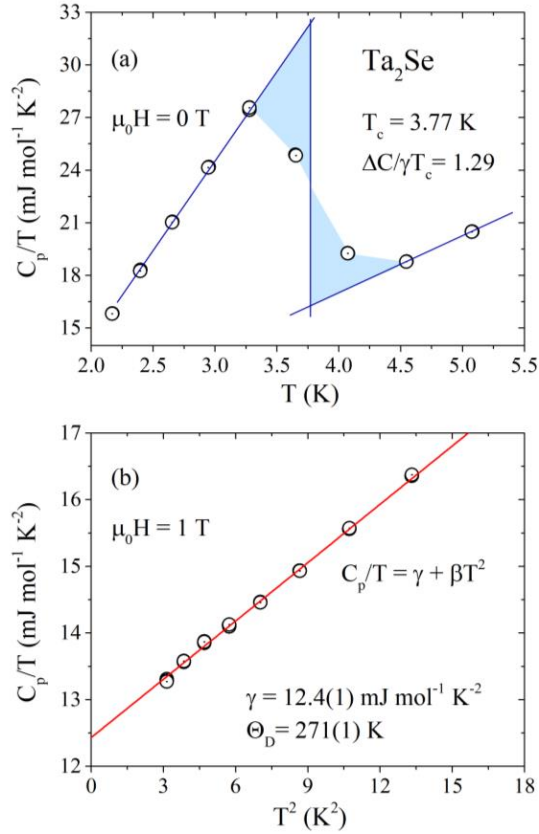

**Figure 3(a).** the specific heat anomaly in zero magnetic field at low temperatures with $T_c = 3.77$ K. **(b).** $C_p/T$ versus $T^2$ plot under a $\mu_0H = 1$ T magnetic field.

Heat capacity measurement by measuring entropy changes during the superconducting transition is a reliable evidence of the presence of bulk superconductivity. To prove that the superconductivity is intrinsic to Ta$_2$Se and is not a consequence of the possible impurity phases in the sample, such as TaSe$_2$ or Ta, the specific heat measurements were conducted on Ta$_2$Se sample. Superconductivity can be considered as a "phase" transition with a superconducting phase transition occurring below the critical temperature. The Figure 3$a$ depicts the closer view of the data under zero magnetic field. Bulk superconductivity was also proven by a significant anomaly at 3.8 K, close to the T$_c$ obtained from resistivity and magnetic measurements. The C$_p$ jumps at T$_c$, estimated by using the equal entropy construction (blue solid lines), is about $\Delta C/T_c$ = 16 mJ mol$^{-1}$K$^{-2}$. Figure 3$b$ illustrates the heat capacity behavior of Ta$_2$Se under external magnetic field of 1 T. The data can be fitted by C$_p$/T = $\gamma$ + $\beta$T$^2$, where $\gamma$ and $\beta$ are determined by electronic and phononic contributions, respectively. The extrapolation gives $\gamma$ = 12.4(1) mJ mol$^{-1}$K$^{-2}$ and $\beta$ = 0.29(1) mJ mol$^{-1}$K$^{-4}$. Furthermore, the Debye temperature can be estimated through the relation $\Theta_D = \left(\frac{12\pi^4}{5\beta}nR\right)^{1/3}$, where R = 8.314 J mol$^{-1}$K$^{-1}$ and n = 3 for Ta$_2$Se. The obtained Debye temperature is 271(1) K, which is larger than the value for pure Ta element ($\Theta_D$= 240 K). Using the Sommerfeld coefficient ($\gamma$ = 12.4(1) mJ mol$^{-1}$K$^{-2}$) and afore-derived specific heat jump at T$_c$, the superconducting parameter $\Delta C/\gamma T_c$ = 1.29 can be calculated. The obtained value is slightly lower than the theoretical value based on the Bardeen–Cooper–Schrieffer (BCS) theory ($\Delta C/\gamma T_c$ ~1.43), likely caused by presence of impurity phases, which is consistent with the powder X-ray diffraction refinement.

With Debye temperature available, the electron-phonon constant $\lambda_{e-p}$ can be obtained through the inverted McMillan equation[26]:

$$\lambda_{e-p} = \frac{1.04 + \mu^* ln\,(\Theta_D/1.45\,T_c)}{(1 - 0.62\mu^*)ln\,(\Theta_D/1.45\,T_c) - 1.04}$$

where $\mu^*$ is the repulsive screened Coulomb part, the value of $\mu^*$ is usually set to 0.13 for intermetallic superconductors. Using $\Theta_D$= 271(1) K and T$_c$ = 3.8 K (obtained from the specific heat measurements), one obtains $\lambda_{e-p}$ = 0.61, which suggests weak electron-phonon coupling behavior.

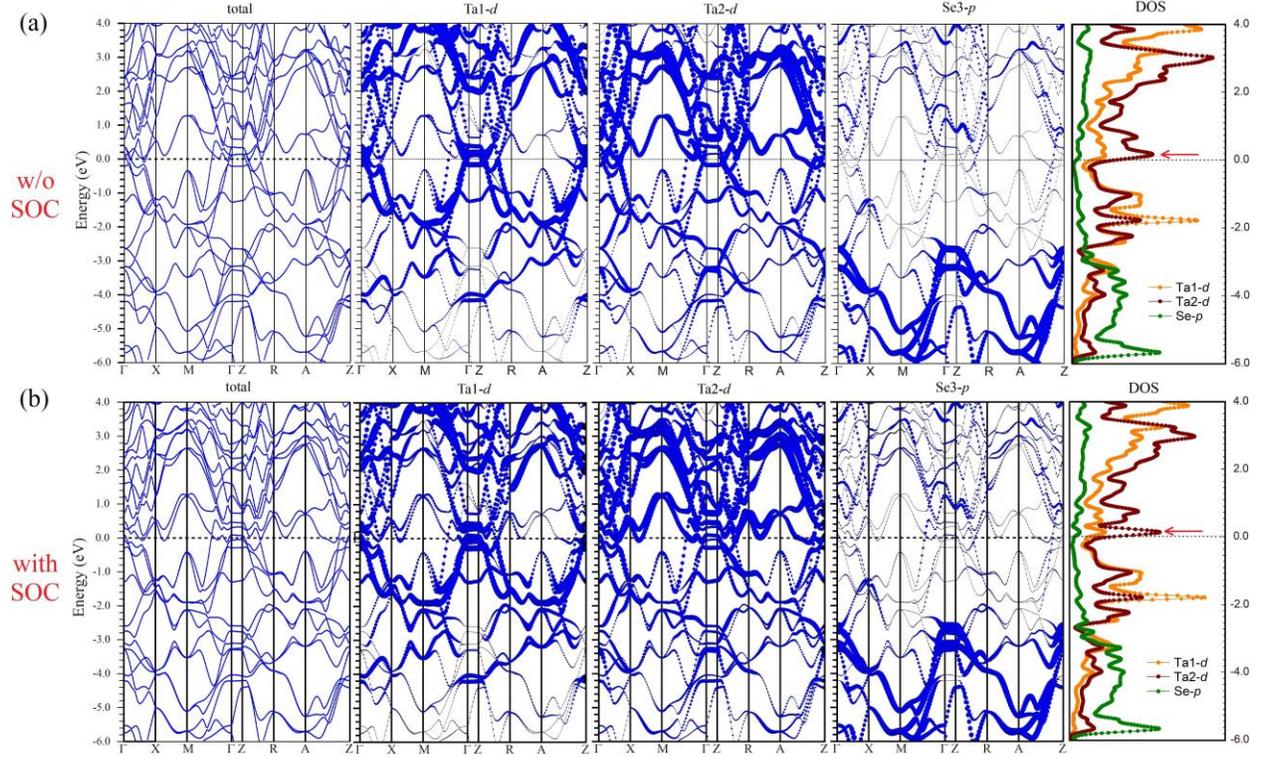

**Figure 4** Band structure of $Ta_2Se$ projected from d orbitals of Ta atoms and p orbitals of Se atom where the thicker the band, the more contribution of the corresponding orbital to the band **(left three)** and density of states of $Ta_2Se$ **(right)** while **(a).** without consideration of SOC effect and **(b).** with consideration of SOC effect.

The total band structure and projection (as shown by the band thickness) of the *d* orbitals in Ta atoms and the *p* orbitals in Se atoms without including the Spin-Orbit Coupling (SOC) effect are calculated, as illustrated in Figure 4. The Fermi levels are dominated by *d* electrons from both Ta sites in both with and without SOC cases, which indicates the critical role of metal-metal bond in stability and superconductivity in $Ta_2Se$. The electrons on *p* orbitals from Se atoms mainly contributes ~ 2.5 eV below Fermi level to stabilize the structure. The flat bands from Γ to Z points near Fermi level, which were mentioned above, are dominated by *d* orbitals of Ta atoms and thus lead to a van Hove singularity near the Fermi level in the DOS. Moreover, after including SOC effect on Ta atoms, a small bandgap (~100 meV) at Γ point ~0.1eV above Fermi level can be observed. The density of states at Fermi level was strengthened by ~10% with including SOC effects marked in the red arrows, ends up with forming a sharp peak at $E_F$. Such intensive DOS at $E_F$ can result in some electronic instability and exotic physical properties such as superconductivity.

Thus, it is highly possible that more metal-rich chalcogenides can host superconductivity due to the dominant metal-metal interaction.

In summary, we report the first metal-rich chalcogenide superconductor $Ta_2Se$ with extensive metal-metal interactions. $Ta_2Se$ was synthesized by the arc-melting method, structurally characterized by powder X-ray diffraction, physical properties measured by magnetism, resistance and specific heat, and discovered to be a new weak-coupling BCS superconductor at Tc ~ 3.8 K. The electronic structures analyzed by WIEN2k program indicate the importance of *d* electrons of Ta atoms in superconductivity and leaves high probability to discover more superconductors in metal-rich chalcogenides.

## *Supporting Information*

The Supporting Information is available free of charge on the ACS Publications website at DOI: xxxxx.

Experimental details; SEM-EDS results; Field-dependence of magnetization.

## *Acknowledgement*

The work at LSU is supported by Arnold and Mabel Beckman Foundation through Beckman Young Investigator (BYI) Award. Work at GUT was supported by National Science Centre (Poland), grant number: UMO-2018/30/M/ST5/00773.

**For Table of Contents Only**

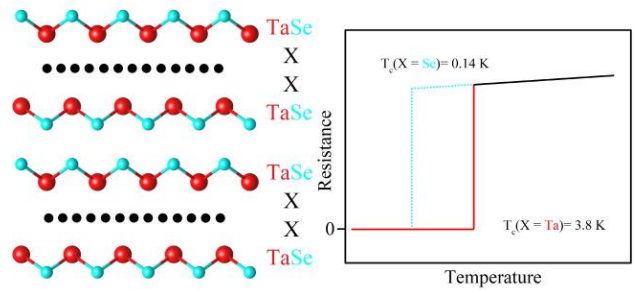

The first metal-rich chalcogenide superconductor Ta$_2$Se with extensive metal-metal interactions was reported with T$_c$ ~ 3.8 K. The electronic structures indicate the importance of *d* electrons of Ta atoms in superconductivity and leaves high probability to discover more superconductors in metal-rich chalcogenides.